\documentclass{article}

\PassOptionsToPackage{numbers, compress}{natbib}

\usepackage[final]{neurips_2019}

\usepackage{graphicx} \graphicspath{ {figures_corners/} } 

\usepackage{subfigure}
\usepackage{amsmath}
\usepackage{amssymb}
\usepackage[utf8]{inputenc} 
\usepackage[T1]{fontenc}    
\usepackage{hyperref}       
\usepackage{url}            
\usepackage{booktabs}       
\usepackage{amsfonts}       
\usepackage{nicefrac}       
\usepackage{microtype}      

\title{Bipartite Distance for Shape-Aware Landmark Detection in Spinal X-Ray Images}

\author{
  Abdullah-Al-Zubaer Imran${}^{2,1}$, \quad Chao Huang${}^1$, \quad Hui Tang${}^1$, \quad Wei Fan${}^1$, \\[2pt]  \textbf{Kenneth M.C. Cheung${}^3$, \quad Michael To${}^3$,  \quad Zhen Qian${}^1$, \quad Demetri Terzopoulos${}^{2,4}$}\\[5pt]
  ${}^1$Tencent Medical AI Lab, Palo Alto, CA, USA\\
  ${}^2$University of California, Los Angeles, CA, USA\\
  ${}^3$The University of Hong Kong, China\\
  ${}^4$VoxelCloud, Inc., Los Angeles, CA, USA
}

\begin{document}

\maketitle

\begin{abstract}
  Scoliosis is a congenital disease that causes lateral curvature in the spine. Its assessment relies on the identification and localization of vertebrae in spinal X-ray images, conventionally via tedious and time-consuming manual radiographic procedures that are prone to subjectivity and observational variability. Reliability can be improved through the automatic detection and localization of spinal landmarks. To guide a CNN in the learning of spinal shape while detecting landmarks in X-ray images, we propose a novel loss based on a bipartite distance (BPD) measure, and show that it consistently improves landmark detection performance.
\end{abstract}

\section{Introduction}

Scoliosis is an abnormal condition characterized by lateral spinal curvature. Early assessment and treatment planning is critical \citep{weinstein2008adolescent}. Conventionally, the assessment of scoliosis is performed manually by clinicians through the identification and localization of vertebral structures in spinal X-ray images. However, large inter-patient anatomical variation and poor image quality challenge clinicians to assess the severity of scoliosis accurately and reliably. Automated measurement promises to enable the reliable quantitative assessment of scoliosis. 

Several spinal landmark detection methods are available in the literature: Conventional hand-crafted feature engineering \citep{ebrahimi2019vertebral} is a semi-automatic method involving several sub-tasks. Our approach is automatic convolutional neural network (CNN) models. The CNN model of Wu et al.~\citep{wu2017automatic} requires cropped images and tedious data augmentation. Landmarks can also be detected by segmenting the relevant vertebrae \citep{imran2019end}. Our proposed model is totally end-to-end, requiring no pre-processing, and is fully automatic, eschewing any hand-crafted feature extractions.

\section{Method}

Given an X-ray image, we formulate the landmark detection problem as identifying $n$ landmarks localizing the relevant vertebrae. Each training image $x_i$, for $i=1,\dots, m$, is annotated by an associated $2n$-dimensional landmark vector $y_i$. Through supervised learning, a CNN can be trained to extract landmarks automatically, by minimizing the standard mean squared error (MSE) loss  
\begin{equation}
    \text{MSE} = \frac{1}{m}\sum_{i=1}^m (y_i - \hat{y}_i)^2,
\end{equation}
where $y_i$ are the ground-truth landmarks and $\hat y_i$ are the predicted landmarks. However, the MSE loss ignores inter-landmark relationships. To guide a CNN in the detection of landmark coordinates while learning spinal shape, we propose a novel distance measure---bipartite distance.

\begin{figure}
    \centering
    \begin{tabular}{cc}
    \includegraphics[width=0.12\linewidth, trim={380 180 470 80}, clip]{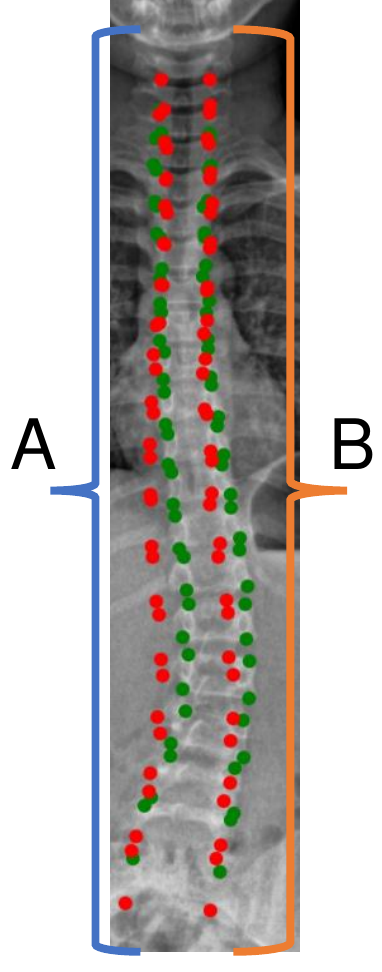}
    &
    \includegraphics[width=0.79\linewidth, trim={0cm 19cm 3cm 2cm}, clip]{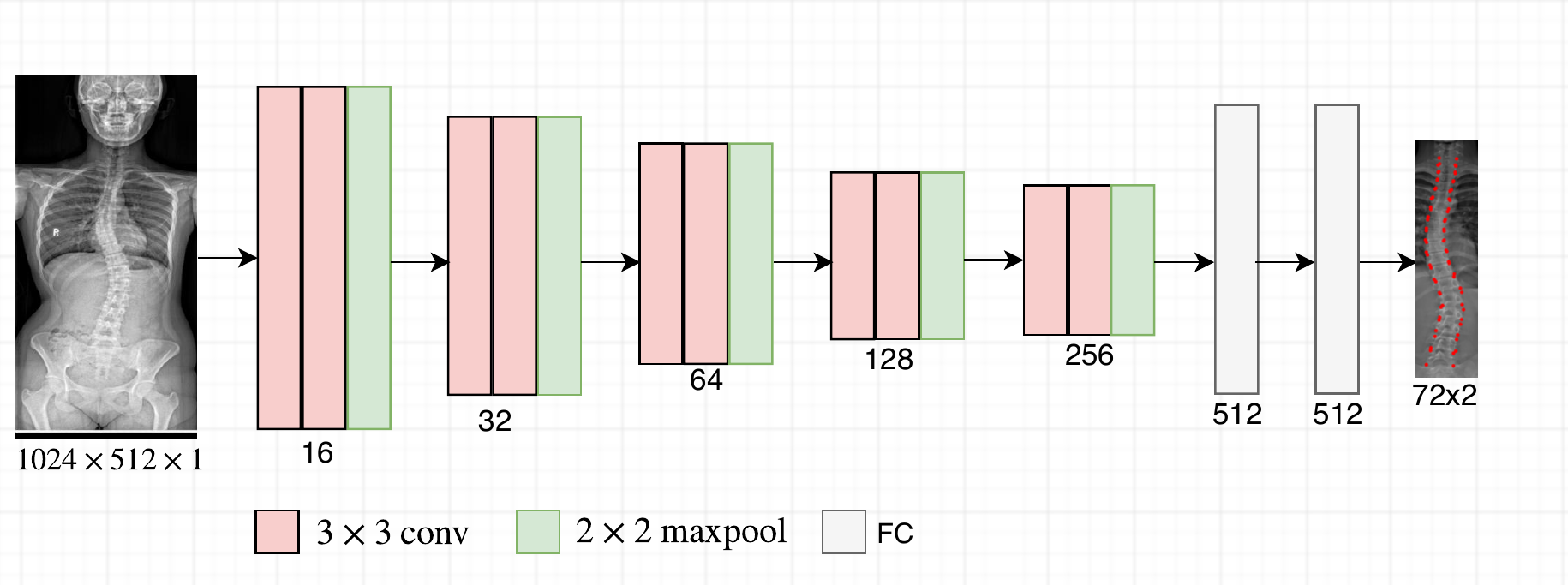}\\
    (a) & (b)
    \end{tabular}
        \caption{Convolutional neural networks for landmark detection from spine X-ray images: (a) Illustration of the bipartite distance in a spinal image based on the ground truth (green) and predicted (red) landmarks. (b) Model architecture.}
    \label{fig:model}
\end{figure}

Referring to Figure~\ref{fig:model}a, we regard the ground-truth (green) landmarks $A_y$ on the left and $B_y$ on the right of the spine as the two disjoint sets of vertices of a complete bipartite graph whose edges connect every landmark in $A_y$ with all landmarks in $B_y$. The same holds for the predicted (red) landmarks, $A_{\hat{y}}$ and $B_{\hat{y}}$. This leads to a shape-aware loss, which penalizes the CNN model when the pairwise distances between the predicted landmarks deviate from those between the ground truth landmarks. Letting $d^e$ denote the Euclidean distance between ground-truth landmarks connected by edge $e$ of the graph and $\hat d^e$ denote the Euclidean distance between the corresponding predicted landmarks, the bipartite distance (BPD) is
\begin{equation}
    \label{eqn:bpd}
    \text{BPD} = \frac{1}{m}\sum_{i=1}^{m} \sum_e\left|d_i^e - \hat d_i^e\right|.
\end{equation}

We employ the loss function
\begin{equation}
    \label{eqn:loss}
    L = \text{MSE} + \alpha\,\text{BPD},
\end{equation}
where $\alpha$ weighs the BPD term against the MSE.

\section{Implementation Details}

Our dataset consists of 100 high-resolution anterior-posterior spinal X-ray images with signs of mild to severe scoliosis. Since the cervical vertebrae are seldom involved in spinal deformity and the identification of the bottom cervical vertebra could be important, we selected 18 vertebrae: C7 (cervical), T1--T12 (thoracic), and L1--L5 (lumbar). Medical experts provided binary segmentation annotation by labeling the $n/4=18$ vertebrae in the X-ray images. The 4 corners of each vertebral region serve as landmarks. They were automatically extracted by applying FAST \citep{rosten2006machine} to the expert-segmented labels. Therefore, associated with each spinal image are 72 landmarks to be estimated. 

As shown in Figure~\ref{fig:model}b, our model is a CNN comprising five convolutional layers and three fully-connected (FC) layers. Leaky-ReLU is used as the activation function in each layer. The convolutional layers have feature sizes 16, 32, 64, 128, and 256. In each layer, two $3\times 3$ convolution operations are followed by a $2\times 2$ maxpooling layer. After every convolutional layer, we use a batch-normalization layer and a dropout layer with the rate of 0.25.  After two FC layers with 512 neurons, a final FC layer of $2n=144$ neurons is used to produce the image-plane coordinates of the landmarks. The model is implemented in Tensorflow with Python 3 and runs on a Tesla P40 GPU on a 64-bit Intel(R) Xeon(R) 440G CPU.

The dataset was split into training (80 images), testing (15 images), and validation (5 images) sets. All the images were resized to $1024\times 512\times 1$ and normalized to $[0, 1]$ before feeding them to the network. When the model is trained using our MSE-BPD loss, we used $\alpha=0.01$ in (\ref{eqn:loss}). As a baseline, we trained the same architecture using only the MSE loss; i.e., $\alpha=0$. The models were trained with a minibatch size of 4. We used the Adam optimizer with a learning rate of 0.0001 and momentum 0.9. 

\section{Experimental Results}

\begin{figure}
    \centering
    \resizebox{\linewidth}{!}{
    \begin{tabular}{|ccc|ccc|}
    {\Large Raw Image} & {\Large MSE} & {\Large MSE-BPD} & {\Large Raw Image} & {\Large MSE} & {\Large MSE-BPD}\\[10pt]
    \includegraphics[width=0.3\linewidth]{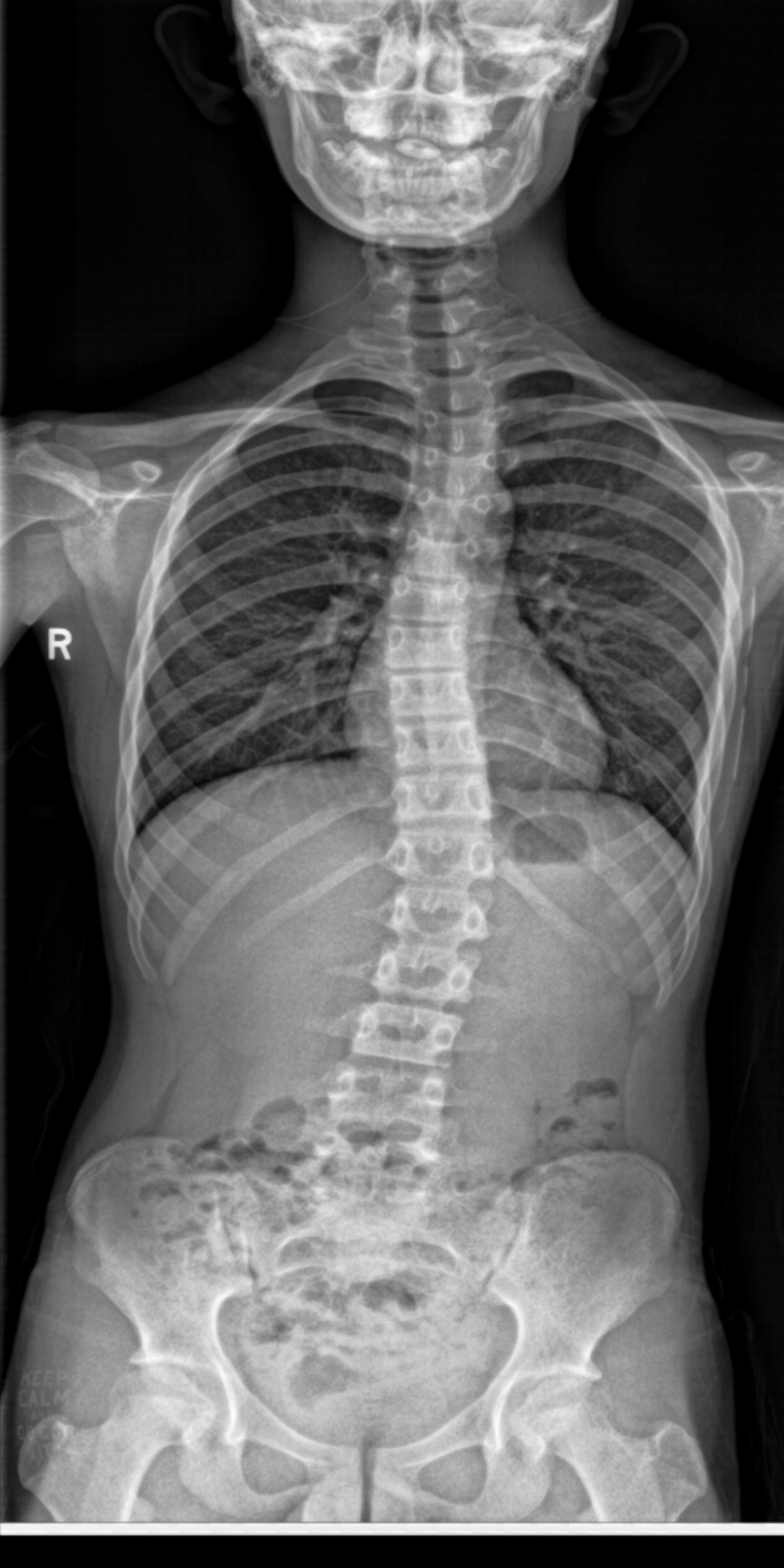}
    &
    \includegraphics[width=0.2\linewidth, trim={12cm 4cm 11cm 4cm}, clip]{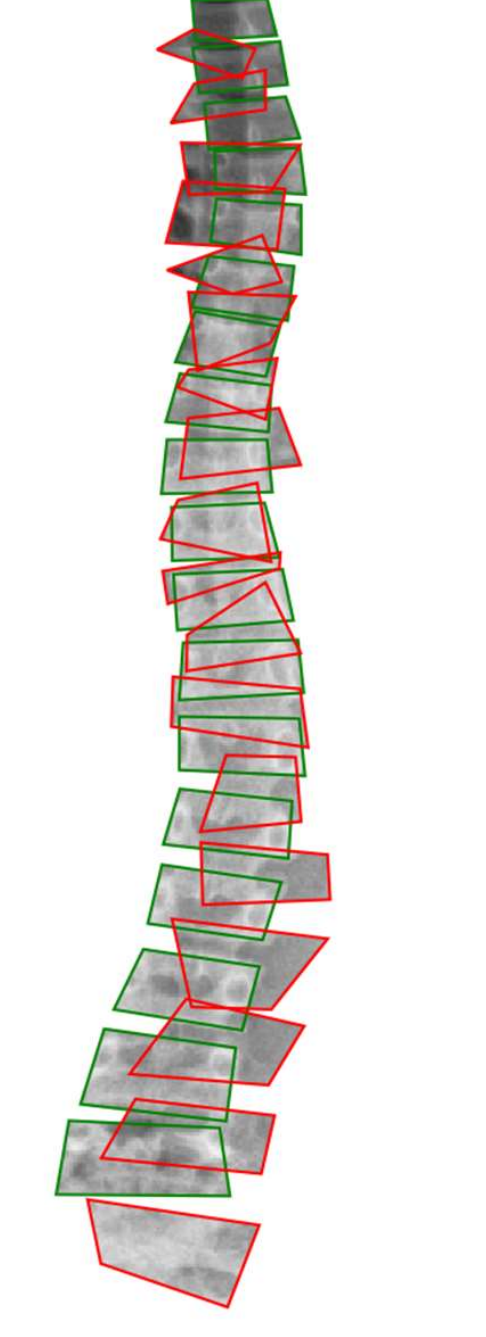}
    &
    \includegraphics[width=0.2\linewidth, trim={12cm 4cm 11cm 4cm}, clip]{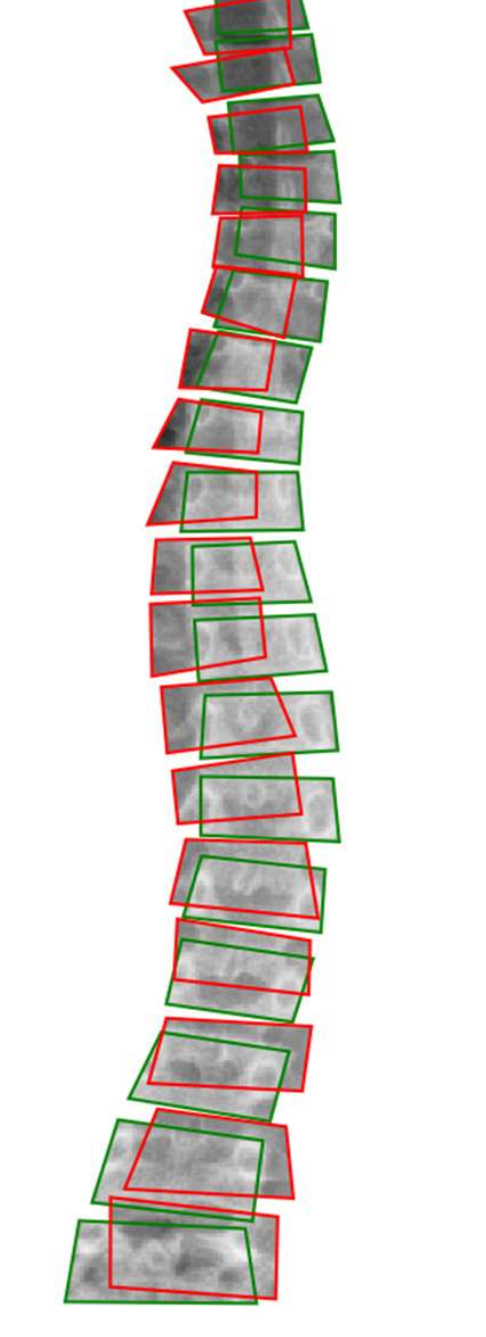}
    &
    \includegraphics[width=0.3\linewidth]{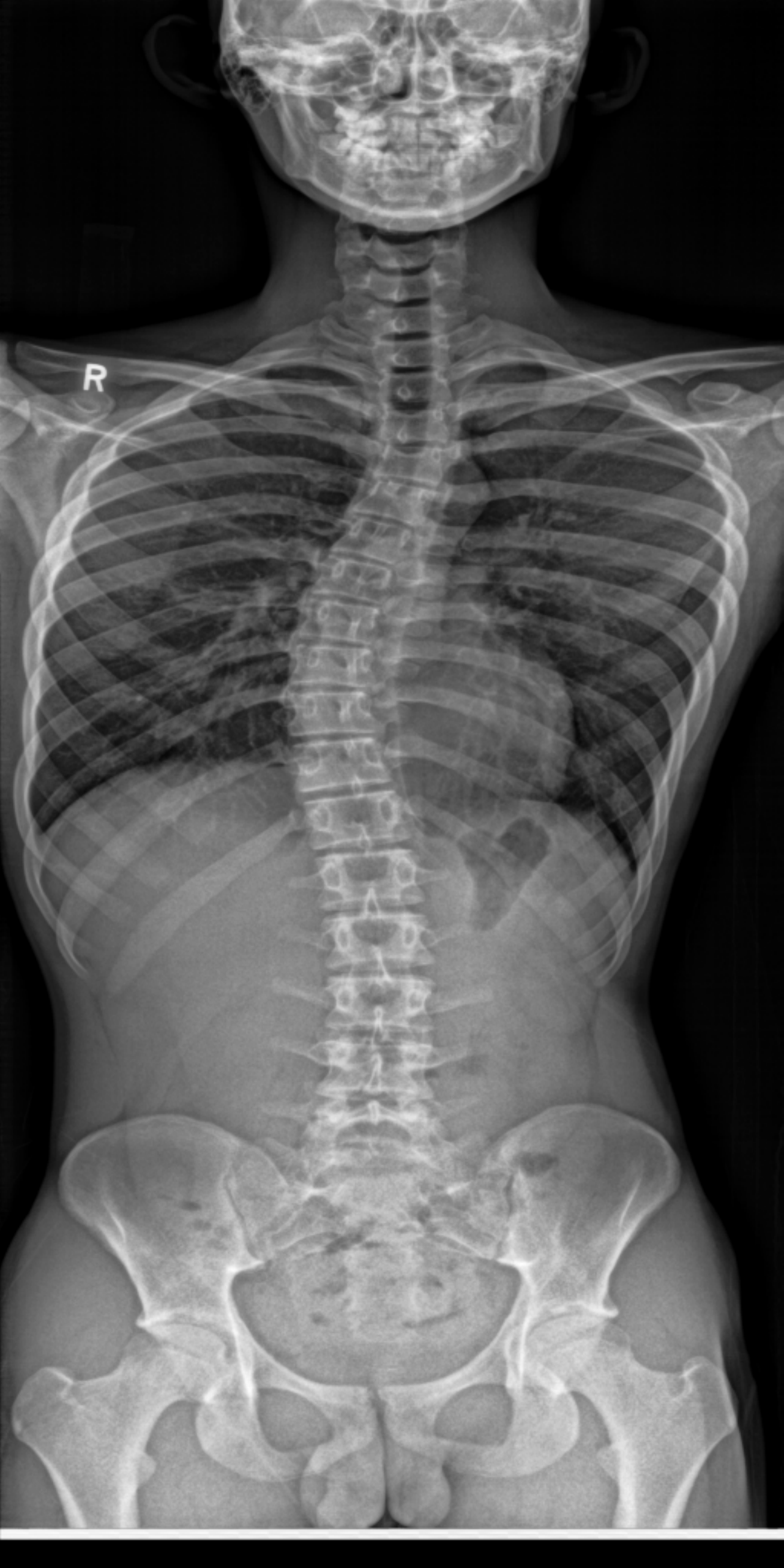}
    &
    \includegraphics[width=0.2\linewidth,trim={12cm 4cm 11cm 4cm}, clip]{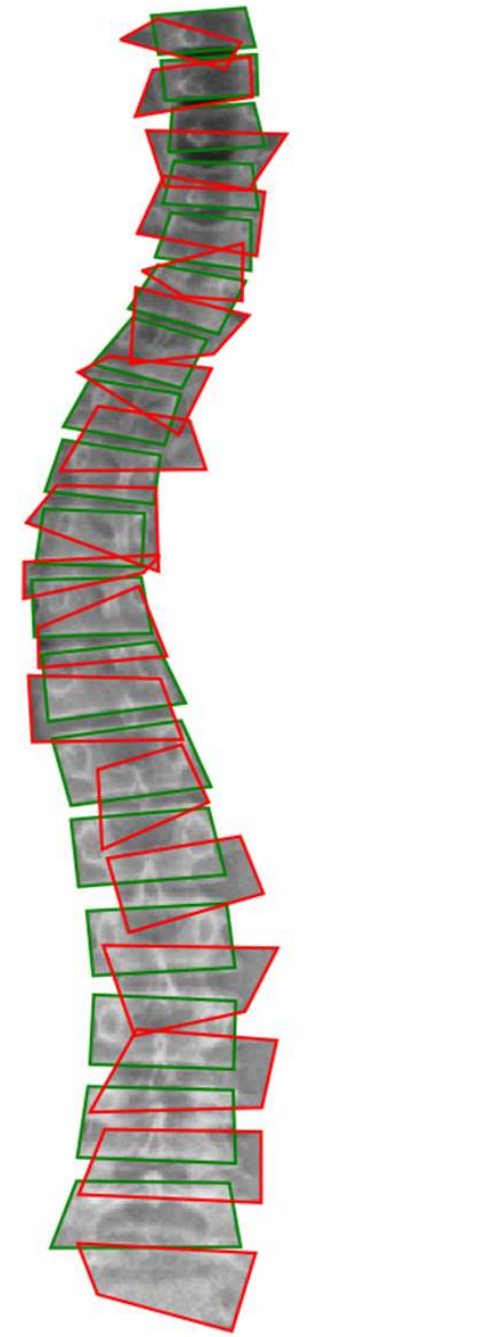}
    &
    \includegraphics[width=0.2\linewidth,trim={12cm 4cm 11cm 4cm}, clip]{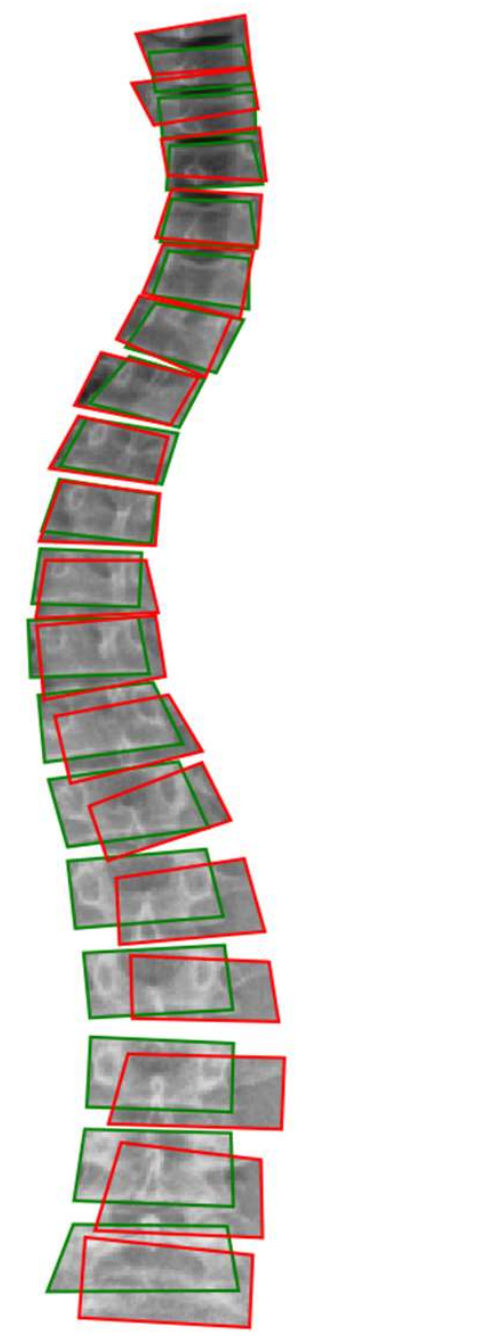}
    \\
    \end{tabular} }
    \caption{Qualitative comparison shows improved spinal shape and landmark detection performance by our model (MSE-BPD loss) relative the baseline model (MSE loss) in two spinal X-ray images from the test set. Green boxes bound vertebrae based on the ground-truth landmarks; red boxes bound vertebra based on the model-predicted landmarks.}
    \label{fig:bpd}
\end{figure}

We compared the performance of the proposed model (MSE-BPD) against the baseline (MSE) both qualitatively and quantitatively. Qualitative comparisons (e.g., Figure~\ref{fig:bpd}), show better agreement of our model over the baseline model. The irregular spinal shapes in the baseline model are mitigated by our model. Moreover, the landmark detection performance is also improved in our model, which achieves a correlation score of 0.95 compared to the baseline model's score of 0.92 (Pearson correlation coefficient). Moreover, one-way ANOVA analysis confirms that the landmarks predicted by our model have no significant difference with the ground truth landmarks ($\textit{p-value} <0.05$). 

\section{Conclusions}

The detection of vertebral landmarks is crucial for the accurate measurement of scoliosis in spinal X-ray images. To this end, we proposed a new loss function which guides the training of a CNN vertebral (corner) landmark detection model to perform reliable shape-aware predictions. 

\bibliographystyle{agsm}
\bibliography{references}
\end{document}